\documentclass{article}

\usepackage{jmlr2e}
\usepackage{natbib}
\usepackage{lineno,hyperref}
\modulolinenumbers[5]

\usepackage{amssymb,amsmath,epsfig}

\usepackage{caption}

\usepackage[utf8]{inputenc}

\usepackage{algorithm}
\usepackage{algorithmic}
\floatname{algorithm}{Algorithm}
\renewcommand{\algorithmicrequire}{\textbf{Input:}}
\renewcommand{\algorithmicensure}{\textbf{Output:}}

\usepackage[french,english]{babel}

\usepackage{color}
\hypersetup{
     colorlinks   = true,
     linkcolor=red,
     urlcolor=green,
     citecolor    = blue
}

\usepackage{picins}
\usepackage{enumitem}
\usepackage{multirow}

\begin{document}

\title{Beyond Voice Identity Conversion: Manipulating Voice Attributes by Adversarial Learning of Structured Disentangled Representations}


\author{\name Laurent Benaroya \email Laurent.Benaroya@ircam.fr\\
       \addr STMS Lab. \\
       Ircam, CNRS, Sorbonne Université\\
       1, place Igor Stravinsky\\
       Paris, France \\
       \name Nicolas Obin \email Nicolas.Obin@ircam.fr\\
       \addr STMS Lab. \\
       Ircam, CNRS, Sorbonne Université\\
       1, place Igor Stravinsky\\
       Paris, France \\
       \name Rafael Ferro \email Rafael.Ferro@ircam.fr\\
       \addr STMS Lab. \\
       Ircam, CNRS, Sorbonne Université\\
       1, place Igor Stravinsky\\
       Paris, France \\
       \name Axel Roebel \email Axel.Roebel@ircam.fr\\
       \addr STMS Lab. \\
       Ircam, CNRS, Sorbonne Université\\
       1, place Igor Stravinsky\\
       Paris, France}


\maketitle

\begin{abstract}
 Voice conversion (VC) consists of digitally altering the voice of an individual to manipulate part of its content, primarily its identity, while maintaining the rest unchanged. Research in neural VC has accomplished considerable breakthroughs with the capacity to falsify a voice identity using a small amount of data with a highly realistic rendering. This paper goes beyond voice identity and presents a neural architecture that allows the manipulation of voice attributes (e.g., gender and age).
Leveraging the latest advances on adversarial learning of structured speech representation, a novel structured neural network is proposed in which multiple auto-encoders are used to encode speech as a set of idealistically independent  linguistic and extra-linguistic representations, which are learned adversariarly and can be manipulated during VC. Moreover, the proposed architecture is time-synchronized so that the original voice timing is preserved during conversion which allows lip-sync applications.
Applied to voice gender conversion on the real-world VCTK dataset, our proposed architecture can learn successfully gender-independent representation and convert the voice gender with a very high efficiency and naturalness. 
\end{abstract}

\begin{keywords}
Voice Conversion, Representation Learning, Adversarial Learning
\end{keywords}


\section{Introduction}
\label{sec:intro}

\subsection{Context}
Voice conversion (VC) consists of digitally altering the voice of an individual - e.g., its identity, accent, or emotion - while maintaining its linguistic content unchanged. Primarily applied to identity conversion \citep{KUWABARA1995165, Sty98}, VC has considerably gained in    popularity and in quality thanks to the advances accomplished with neural VC (see the two editions of the VC challenge for a short review of the latest challenges and contributions: \citep{Tod16, Lor18, zhao2020voice} . 
Similarly to face manipulation, voice conversion has a wide range of potential applications, such as voice cloning and deep fake \citep{lorenzotrueba2018steal} in the fields of entertainment and fraud, anonymization of voice identity \citep{Vincent20, ericsson2020adversarial} in the field of security and data privacy, or digital voice prosthesis of impaired speech \citep{Wan20} in the field of digital healthcare.
In its original formulation, the VC task consisted into learning the one-to-one statistical acoustic mapping between a pair of source and target speakers from a set of shared and temporarily pre-aligned set of utterances \citep{Sty98}. The use of the same sentences shared among speakers and the pre-alignment greatly facilitated the learning since the mapping could directly be learned from this set of perfectly paired data. On the other side, this constraint also greatly limited the size of the dataset that could be used for learning, and subsequently the quality of the obtained conversion. From this first formulation, many advances have been proposed through years including the one-to-many, many-to-one and many-to-many VC in which a set of multiple speakers is used as prior knowledge to pre-trained conversion functions which are further adapted to an unseen utterance or speaker during conversion \citep{Toda07}. 

Neural VC, (i.e. VC based on neural networks) has been first introduced by \citep{desai_voice_2009}, following the one-to-one and parallel VC paradigm and by simply replacing GMM by NN in order to estimate the conversion function. Through the considerable advances established in the theory and applications of neural networks in computer vision, image processing, and natural language processing, among which: Generative Adversarial Networks \citep{goodfellow_generative_2014} (GAN), Sequence-to-Sequence Models \citep{sutskever2014sequence, BahdanauCB14} (S2S), Variational Auto-Encoders \citep{hsu_voice_2016} (VAE), cycle-GAN \citep{zhu_unpaired_2017}, star-GAN \citep{Choi18}, and disentanglement learning \citep{higgins2018towards}. Neural VC  is now widely considered as a standard in VC and has reached a highly-realistic rendering of voice identity conversion from a small amount of data of a target voice.

\subsection{Related works and contributions}

\begin{figure}[h!]
\begin{center}
\includegraphics[width=0.7\columnwidth]{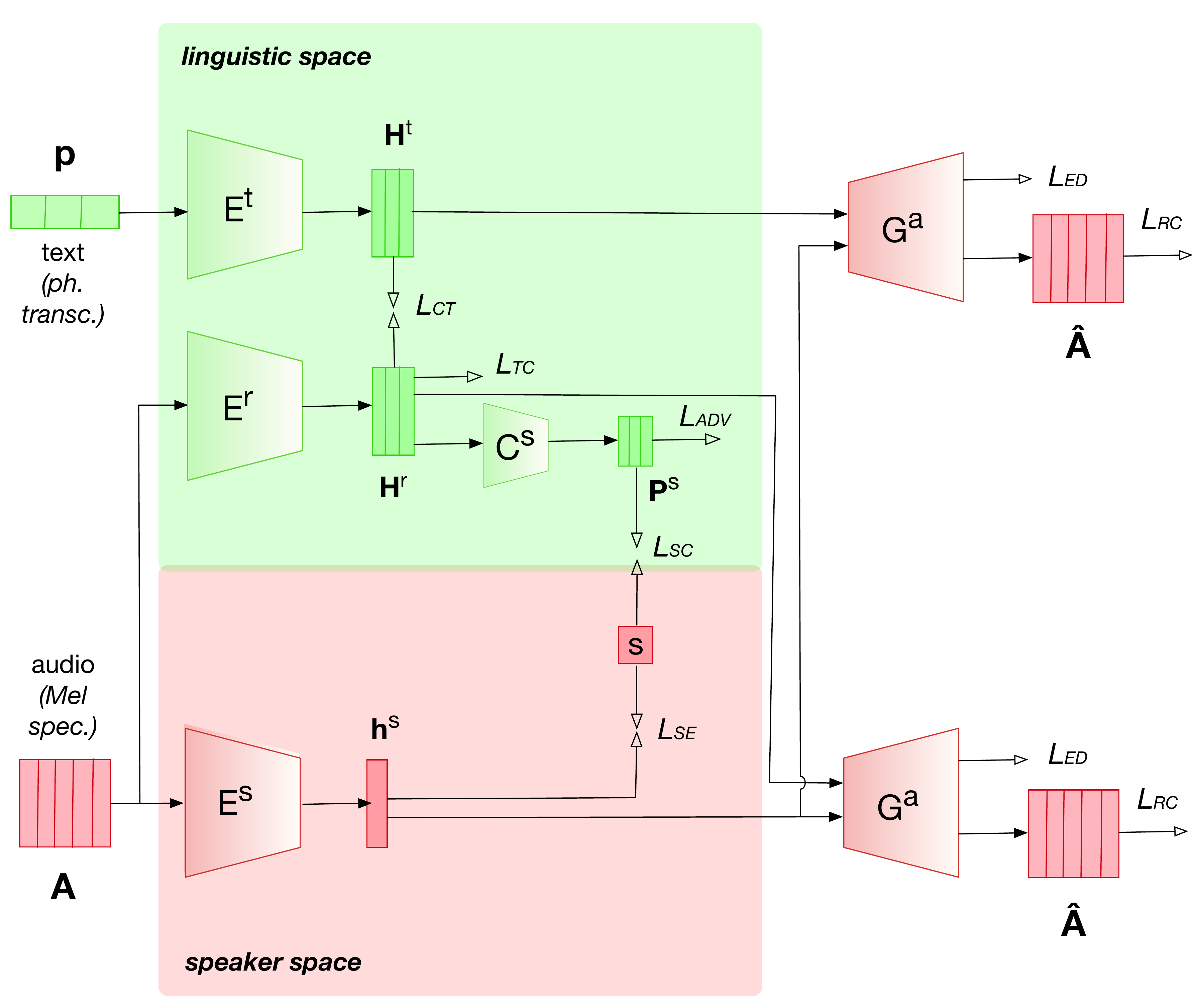}
{\hspace{0.4cm}
\includegraphics[width=0.7\columnwidth]{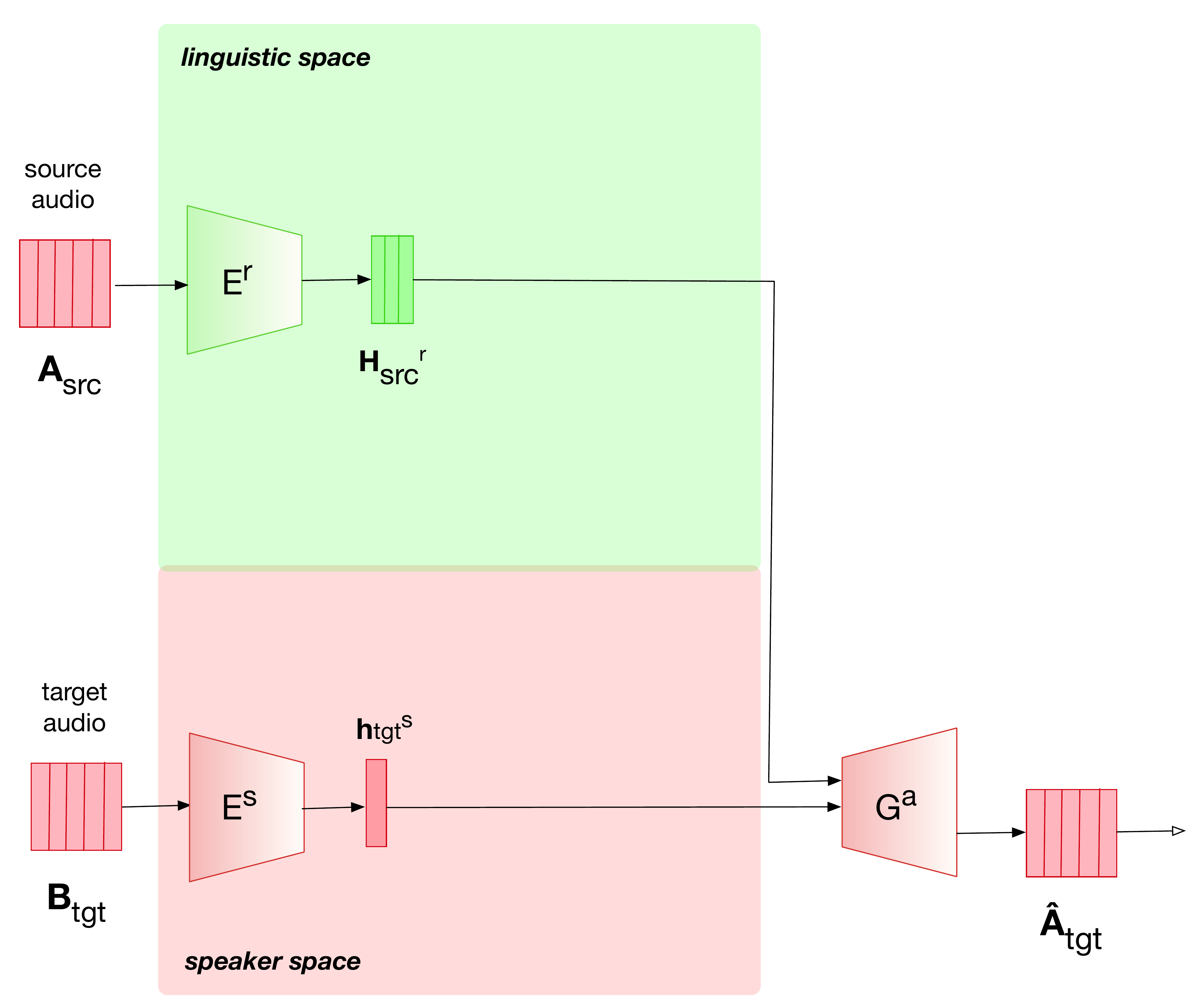}}
\caption{Architecture of the original S2S VC system with adversarial learning of disentangled linguistic and speaker representation. On left: training phase. On right: conversion phase. See Section \ref{sec:S2SVC_orig} for description.} 
\label{figure:S2SVC_orig}
\end{center}
\end{figure}

Through the multiple and various contributions in neural VC, one can distinguish a progressive but strong evolution in the VC paradigm: from the initially agnostic learning of one-to-one VC using parallel datasets to today's structured and informed learning of many-to-many VC from non-parallel datasets. 

Following the historical paradigm of one-to-one parallel VC, cycle-GAN  and S2S with attention mechanisms.   VC models have been first proposed to learn the acoustic mapping from pairs of sentences of source and target speakers. As inspired by \citep{zhu_unpaired_2017}, cycle-GAN VC \citep{ kaneko_parallel-data-free_2017, kaneko_cyclegan-vc2:_2019, fang_high-quality_2018} attempts to learn the identity conversion function in both directions though a cycle. Beyond the usual GAN losses, the cycle-consistency loss is assumed to stabilize the learning by encouraging to preserve the linguistic content (seen as a ‘‘background'') unchanged during conversion. In the S2S VC \citep{tanaka_atts2s-vc:_2018, kameoka_convs2s-vc:_2018}, the conversion is formulated in the form of a recurrent encoder and decoder, at the interface of which an attention mechanism \citep{BahdanauCB14}  is used to learn the alignment between the recurrent encodings of the source and target speakers sequences, thus optimizing the sequential learning of the conversion. 
However, the one-to-one VC framework using parallel dataset remains highly limited: the size of the parallel dataset is too small to learn efficiently a conversion, and there is no solution to exploit knowledge from large and non-parallel dataset to overcome this limitation.

To overcome the shortcomings and limitations of this paradigm, research effort has gradually moved towards many-to-many and non-parallel datasets, allowing the scalability of neural VC to large and multiple speakers datasets, with the assumption that the increase of data will induce a substantial increase in terms of quality and naturalness of the VC. Among the first attempts, starGAN VC \citep{kameoka_stargan-vc:_2018, kaneko_stargan-vc2:_2019} has been proposed to extend the paradigm of cycle-GAN to many-to-many and non-parallel VC by proposing a conditional encoder-decoder architecture. As opposed to the cycleGAN VC, starGAN VC is composed of a single encoder-decoder in which the decoder is conditioned on the speaker identity attribute to be reconstructed. In addition to the usual cycle-consistency and discriminator losses of a cycle-GAN, a classifier loss is added to determine the speaker identity from the converted speech. 
Further research attempted to break the need to learn any conversion function (either one-to-one or many-to-many) by formulating the VC problem as a conditional auto-encoder \citep{zhou_voice_2018, Lu2019, qian2019autovc}. Similarly to the starGAN, the architecture is an auto-encoder in which the encoder part encodes the source speaker from the input source speaker's utterance, and the decoder part reconstructs the target speaker's utterance from the source encoding and a speaker embedding. The fundamental difference lies in the fact that during training, the source and the target speakers are simply the same. During conversion, one only needs to manipulate the speaker attribute in the decoder to convert the input speech to the desired target identity. This breakthrough has opened the way to VC from a very small number of examples of the target speaker (at the extreme from one-shot \citep{Lu2019} or zero-shot \citep{qian2019autovc}).

In a trade-off between data formatting and model simplicity, the possibility to learn VC from unconstrained data raises a need for more structured neural networks. In particular, latest research tend to integrate prior linguistic knowledge about speech communication to construct structured architectures that aims to learn and manipulate those speech information independently. In the theory of speech communication, speech conveys three types of information: linguistic (the primary meaning, i.e., the text), para-linguistic (secondary information that helps to understand the intended meaning, e.g., modality of a question or emotional state of the speaker), and extra-linguistic (which provides only information about the speaker, e.g., speaker identity, socio-geographical origins). Some late VC architectures are structured so as to integrate explicitly prior knowledge about linguistic content and speaker identity \citep{Zhang19}, e.g., by the explicit use of textual information (Phonetic Posterior-Grams, PPG \citep{sun_phonetic_2016, mohammadi2019one}) and with speaker representation inherited from the domain of speaker authentication \citep{SpeakerVerificationTTS}. 

Through bottleneck or disentangled representations \citep{fader2017, higgins2018towards}, it has become possible to learn more structured representations: in \citep{qian20a}, three bottlenecks are used to encode separately speech parameters such as pitch, timbre, and rhythm; in  \citep{Zhang_2020, yuan2021improving}, adversarial learning of disentangled representations is employed to learn a set of representations that encodes specifically linguistic information and speaker identity, and idealistically independent from each others. These latest VC systems can achieve highly-realistic voice identity conversion from less and less data of the target speaker, with a similarity and a naturalness of the converted speech that are comparable to those of the original voice. While the VC identity task may appear now somehow solved, it opens new frontiers and new challenges for more advanced and sophisticated voice manipulation beyond voice identity. In particular, the existing solutions to voice identity conversion are extremely specific to voice identity and can not be applied straightforwardly to other attributes, such as age or gender (as this is a common case in image with face manipulation \citep{fader2017, Choi18}). Based on those aforementioned observations and leveraging the disentangled neural VC architecture presented in \citep{Zhang_2020} and inspired by disentanglement of speaker representations \citep{noe2020adversarial}, this paper proposes a structured neural VC architecture for the manipulation of voice attributes. The main contributions of the paper are:
\begin{itemize}[label=$-$]
    \item a neural network framework that allows  manipulating selected  voice attributes during VC (linguistic and extra-linguistic). Building upon a VC system that represents speech in disentangled speaker and linguistic spaces \citep{Zhang_2020} we here propose a new method that allows manipulating individual attributes of the speaker identity. To do so the speaker identity code is disentangled further into a gender-independent identity code and a gender attribute. While in the present study the framework is applied to the task of gender manipulation, we foresee further extending it to manipulate other identity related attributes such as age, speaking style, or attitudes;
    \item A time-synchronized VC algorithm ensuring that the converted speech is temporarily synchronized with the source speech. This is obtained by exploiting phoneme duration and preserving them all along the VC network, by proposing a formulation which can both encode time-dependent and time-independent information. The time-synchronization allows the integration of the proposed VC algorithm to audio-visual face manipulation \citep{Thies18} or using it for cinema productions, which require that the converted speech remains synchronized with the lip movements in the video. 
\end{itemize}

The remaining of the paper is organized as follows: Section \ref{sec:contribution} presents the original S2S VC system which serves as the reference VC architecture. This is followed by a description the two proposed contributions: time-synchronized VC and disentanglement of speaker identity and attribute during VC. Section \ref{sec:experiments} presents a complete experimental evaluation of the proposed VC architecture with application to voice gender manipulation including objective and subjective experiments.

\section{Neural VC: Manipulating Identity and Beyond}
\label{sec:contribution}

\subsection{Original S2S Neural VC}
\label{sec:S2SVC_orig}

The original non-parallel VC system presented in \citep{Zhang_2020} has a form of a sequence-to-sequence (S2S) auto-encoder in which contrastive and adversarial losses are introduced to disentangle speaker identity and linguistic representations. The architecture of the S2S-VC system is presented in Figure \ref{figure:S2SVC_orig}. The inputs of the system are the sequence of $N$ phonemes $\mathbf{p}$ corresponding to the phonetic transcription of the input text and the audio signal matrix $\mathbf{A}$ represented the Mel-spectrogram computed on $T$ time frames. Dual encoders $E^t$, $E^r$, and $E^s$ are employed to encode linguistic and speaker information. The speaker encoder $E^s$ converts the audio signal $\mathbf{A}$ into a time-independent vector $\mathbf{h}^s$, since it is assumed that the identity of a speaker is not dependent on time through an utterance. A speaker classification loss $L_{SE}$ is defined between the speaker identity predicted from $\mathbf{h}^s$ and the true speaker identity $\mathbf{s}$.

During training, two linguistic encoders are used: a text encoder $E^t$ which converts the phoneme sequence $\mathbf{p}$ into a linguistic embedding $\mathbf{H}^t$ which has the same length $N$ as the phoneme sequence; a recognition encoder $E^r$ which converts the audio signal $\mathbf{A}$ in to a linguistic embedding $\mathbf{H}^r$ which has the same length $N$ as the phoneme sequence. During training, a contrastive loss $L_{CT}$ is defined between the linguistic embeddings coming from the text and audio modalities $\mathbf{H}^t$ and $\mathbf{H}^r$ so that these embeddings should be as close as possible at the same phoneme index and dissimilar otherwise. During conversion, the text encoder $E^t$ is discarded and only the recognition encoder $E^r$ is used for conversion. The idea behind this linguistic encoding is to provide a phonetic decoding from the audio signal $\mathbf{A}$. To do so, a text loss $L_{TC}$ is defined as the cross-entropy between the phoneme sequence decoded from $E^r$ and the true phoneme sequence $\mathbf{p}$. 

To disentangle the linguistic and the speaker representation $\mathbf{h}^s$ and $\mathbf{H}^r$, a fader network is introduced to learn them adversiarily. An auxiliary classifier $C^s$ tries to determine the speaker identity from the linguistic embedding $\mathbf{H}^r$, converting this embedding into a matrix $\mathbf{p}^s$ containing the determined speaker identity at each phoneme index. Then, the adversarial learning of disentagled linguistic and speaker representations employs a speaker classification loss $L_{SC}$ defined as a cross-entropy between the speaker identity predicted from $\mathbf{p}^s$ and the true speaker identity $\mathbf{s}$, and an adversarial loss $L_{ADV}$ which is employed to make the latent representation $\mathbf{p}^s$ independent on the speaker identity.

A S2S decoder $G^a$ conditioned on the linguistic embedding $\mathbf{H}^r$ (alt. $\mathbf{H}^t$) and the speaker embedding $\mathbf{h}^s$ is employed to reconstruct an approximation $\widehat{\mathbf{A}}$ of the original audio signal $\mathbf{A}$. During training, two modes are used: a TTS mode which synthesizes the audio signal from the speaker embedding $\mathbf{h}^s$ and the linguistic embedding $\mathbf{H}^t$ calculated from the text transcription, and a VC mode which synthesizes the audio signal from the speaker embedding $\mathbf{h}^s$ and the linguistic embedding $\mathbf{H}^r$ calculated from the audio signal. The decoder $G^a$ is shared among these two modalities. A reconstruction loss $L_{RC}$ is defined between the reconstructed audio signal $\widehat{\mathbf{A}}$ and the original audio signal ${\mathbf{A}}$, and a end-of-decoding loss $L_{ED}$ is defined so as to predict whether the current frame of the decoder is the last frame of the utterance.

During training (see left side of Figure \ref{figure:S2SVC_orig}), the parameters of this S2S VC neural network are trained on multiple-speakers dataset. During the conversion of an utterance from a source speaker to a target speaker, the architecture is greatly simplified (see right side of Figure \ref{figure:S2SVC_orig}): on one hand, the recognition encoder $E^r$ computes the linguistic embedding  $\mathbf{H}^t_{src}$ corresponding to one utterance ${\mathbf{A}_{src}}$ of the source speaker; on the other hand, the speaker encoder $E^s$ computes the speaker embedding $\mathbf{h}^s_{tgt}$ corresponding the one utterance ${\mathbf{B}_{tgt}}$ of the target speaker. Then, decoder $G_a$ is conditioned on the linguistic embedding $\mathbf{H}^t_{src}$ and the speaker embedding $\mathbf{h}^s_{tgt}$ to synthesize the utterance ${\mathbf{A}_{tgt}}$ with the voice of the target speaker.

\subsection{S2S Neural VC for voice attribute manipulation}

\subsubsection{Contribution 1: time-synchronized S2S VC}

\begin{figure}[ht!]
\begin{center}
\includegraphics[width=0.8\columnwidth]{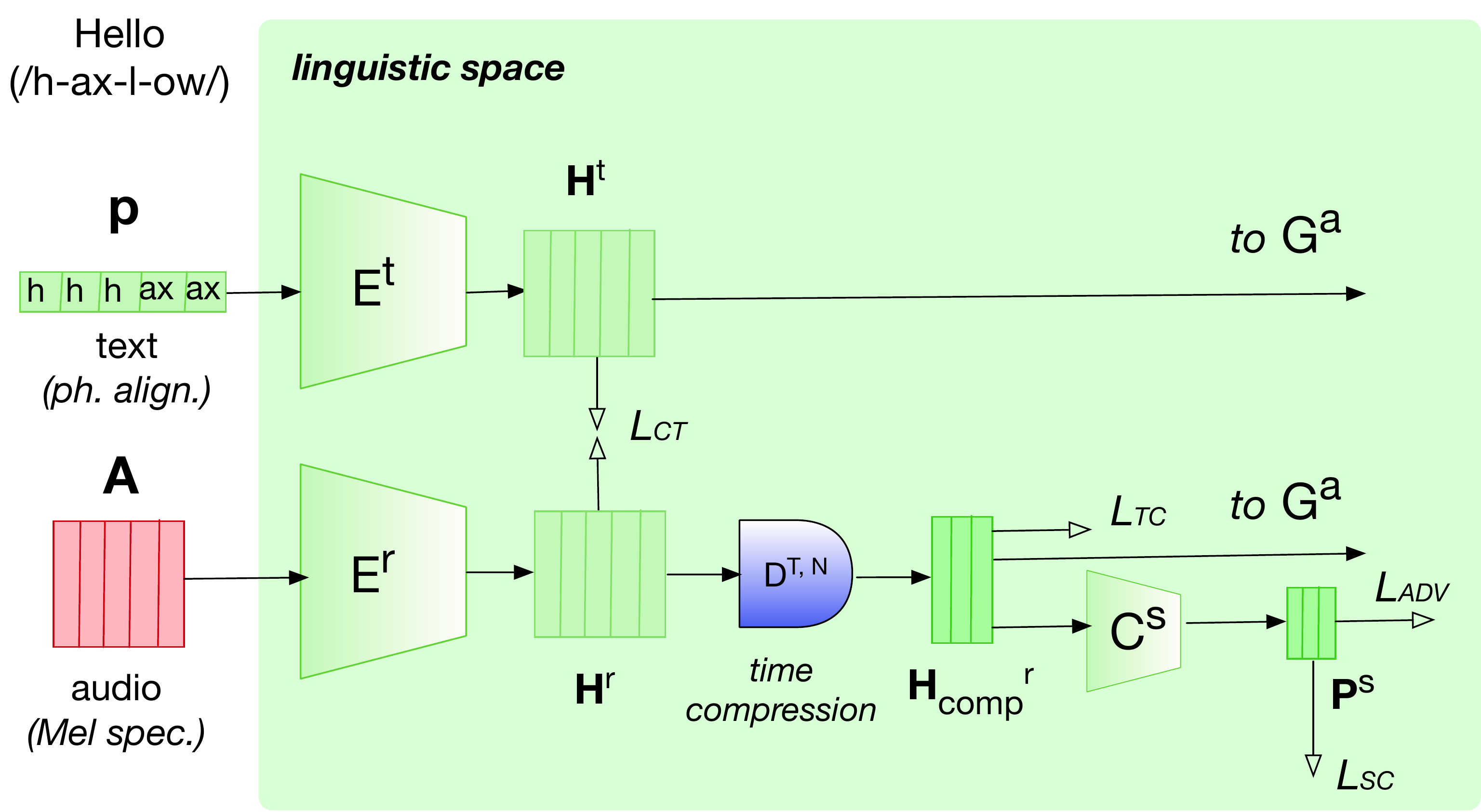}
\caption{Architecture of the proposed time-synchronized S2S VC. Time-synchronization means that the converted speech will be temporally aligned to the original speech signal. For simplicity, only the linguistic space of the architecture is presented.}
\label{figure:S2SVC_timesync}
\end{center}
\end{figure}

In \citep{Zhang_2020}, the $E^r$ recognition encoder makes a temporal compression operation from the $T$ time frames of the audio signal $\mathbf{A}$ to the $N$ phoneme indexes of the linguistic embedding matrix $\mathbf{H}^r$ using an auto-regressive S2S model. In the same manner, the decoder $G^a$ makes a  decompression operation from $N$ phoneme indexes of the linguistic embedding matrix $\mathbf{H}^r$ (resp. $\mathbf{H}^t$) to the $T^\prime$ frames of the reconstructed audio signal $\widehat{\mathbf{A}}$. The decoder has a similar structure to the Tacotron, used for speech synthesis \citep{tacotron2017}, \citep{tacotron2_2018}. 
This allows to make text-to-speech if one uses the text encoder $E^t$ and the decoder $G^a$. The problem is that $T \neq T^\prime $ in general, so that not only the original and converted speech signals do not have the same length, but the two signals will not be temporally synchronized at each time frame. This property prevents the application of this VC architecture for lip-synced audio-visual manipulation, in which the speech timing must be preserved during conversion.

In our proposed framework illustrated in Figure \ref{figure:S2SVC_timesync}, the temporal dimension of the linguistic embeddings $\mathbf{H}^t$ and $\mathbf{H}^r$ are set to the time dimension $T$ of the original audio signal $\mathbf{A}$ (instead of the length $N$ of the sequence of phonemes in the original formulation) and the auto-regressive S2S part of the recognition encoder $E^r$ and the  decoder $G^a$ are modified by incorporating time-synchronous architectures. 
The encoder $E^r$, the text encoder $E^t$ and the decoder $G^a$ operate at a shared temporality which is synchronized to the time of the original audio signal, so that the input audio signal $\mathbf{A}$ and the output of the conversion $\widehat{\mathbf{A}}$ are time-synchronous.
To make the approach consistent with the text input, a phonetic alignment (transcription with timestamps) is computed using a speech recognition with forced alignment. This provides an additional explicit information of the time alignment of each phoneme corresponding to the phonetic transcription of the input text with the audio signal $\mathbf{A}$.
Accordingly, a decompression of the sequence of phonemes is first performed by mapping the original sequence of $N$ phonemes onto the sequence of $T$ time frames of the audio signal $\mathbf{A}$. 
The decompression matrix $D \in \mathcal{R}^{N \times T}$ is defined by~:
\begin{equation}
    D_{n, t} = \mathbb{1}_{\alpha_n = \alpha_t}
\end{equation}
where $\alpha_n$ is the n-th phoneme in the sentence and $\alpha_t$ is the phoneme at frame t. The decompression operation is applied to the input sequence of $N$ phonemes before the text encoder $E^t$ and produces a sequence of $T$ time frames in which the phonemes are replicated according to their duration. The linguistic embeddings $\mathbf{H}^t$ and $\mathbf{H}^r$ then shares the same temporal dimension $T$ as the original audio signal matrix $\mathbf{A}$, from which the contrastive loss $L_{CT}$ is computed.
An opposite compression operation is applied on the linguistic embedding $\mathbf{H}^r$ to compute the phonemes classification loss $L_{TC}$ of the recognition encoder $E^r$, and the adversarial loss $L_{ADV}$ and the classification loss $L_{SC}$ of the speaker classifier $C^s$ of the fader component.
In \citep{Zhang_2020}, the authors employ a pseudo constrastive loss in order to make the linguistic embeddings $\mathbf{H}^r$ of $E^r$ and $\mathbf{H}^t$ of $E^t$ as close as possible for the same time frame and as different as possible for different time frames~: 
\begin{equation}
    L_{CT} = \sum_{t, t^\prime}^{T, T} \left[\mathbb{1}_{t = t^\prime} d_{t, t^\prime} + \mathbb{1}_{t \neq t^\prime} \max(m - d_{t, t^\prime}, 0) \right]
\end{equation}
With:
\begin{equation}
    d_{t, t^\prime} = \left\| \frac{h^r_t}{\|h^r_t\|_2} - \frac{h^t_{t^\prime}}{\|h^t_{t^\prime}\|_2} \right\|_2^2
\end{equation}
where $h^r_t$ comes from $H^r$ at time frame $t$ and $h^t_{t^\prime}$ comes from $H^t$ at time frame $t^\prime$. $m$ is an hyperparameter set to $1.$ in the implementation.

We propose to implement a true contrastive loss as defined in \citep{contrastive_Hadsell_2006}.  Using the compression operation, we can obtain a matrix of distances between the phonemes of the alphabet, i.e. a $P \times P$ matrix where $P$ is the number of symbols in the phonetic alphabet, bringing closer the vectors encoded in $E^t$ and $E^r$ corresponding to same phoneme  and moving them away if they correspond to different phonemes~: 
\begin{equation}
    L_{CT} = \sum_{\alpha, \beta}^{P, P} \left[\mathbb{1}_{\alpha = \beta} d_{\alpha, \beta} + \mathbb{1}_{\alpha \neq \beta} \max(m - d_{\alpha, \beta}, 0) \right]
\end{equation}
where $\alpha$ and $\beta$ are phonemes. 
This approach is more consistent from a theoretical point of view, but does not yield significantly different or better conversion results.
\subsubsection{Contribution 2: S2S VC with voice attribute manipulation}

\begin{figure}[ht!]
\begin{center}
\includegraphics[width=0.8\columnwidth]{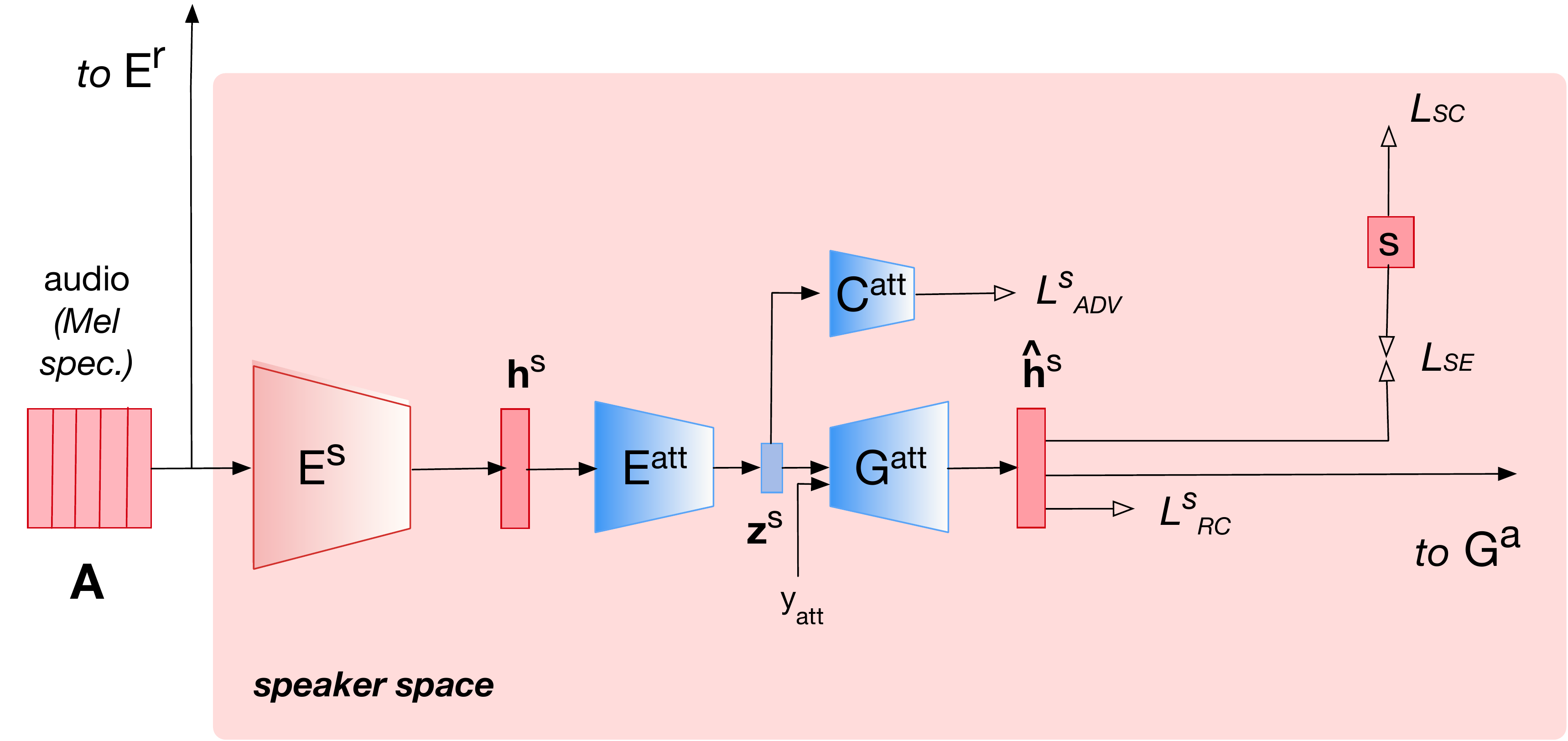}
\caption{Architecture of the proposed speaker disentanglement. The speaker code ${\mathbf h}^s$ is disentangled into an attribute code $att$ and a speaker code ${\mathbf z}^s$ that is independent on the attribute $att$. For simplicity, only the speaker space of the architecture is presented.}  
\label{figure:S2SVC_fader}
\end{center}
\end{figure}

To learn VC that goes beyond voice identity conversion and can be conditioned on other voice attributes, speaker embedding $\mathbf{h}^s$ are further adversiarily disentangled as illustrated in Figure \ref{figure:S2SVC_fader} and inspired by the speaker representations disentanglement described in \citep{noe2020adversarial} in the context of speaker authentication.
In this paper, the proposed speaker disentanglement has a form of a fader network \citep{fader2017}.
The speaker embedding  $\mathbf{h}^s$ resulting from the speaker encoder $E^s$ in the speaker space served as input of the proposed fader network.
This fader network is an autoencoder in which the speaker embedding is encoded by $E^{att}$ to a low-dimensional latent code $\mathbf{z}^s$. 
Conversely, the decoder $G^{att}$ tries to reconstruct the speaker embedding $\widehat{\mathbf{h}}^s$ from the latent code $\mathbf{z}^s$ and the conditioning attribute variable $y_{att}$. In this paper, $y_{att}$ encodes the gender of the speaker as follows: $y_{att} = 0.0$ for female and $y_{att}=1.0$ for male. Additionally, the attribute discriminator $C^{att}$ tries to predict the attribute $y_{att}$ from the latent code $\mathbf{z}^s$
Fader losses are used to make the latent code $\mathbf{z}^s$ independent of the attribute variable $y_{att}$.

The proposed fader network has three losses~: The reconstruction error $L^S_{RC}$, and the two losses for the fader $L^S_{ACC}$ and $L^S_{ADV}$.
The autoencoder loss $L^S_{RC}$ is defined as the mean absolute error between the input speaker embedding $\mathbf{h}^s$ and the reconstructed output speaker embedding $\widehat{\mathbf{h}}^s$,
\begin{eqnarray}
\label{eq:Lrecons}
L^S_{RC} = \| h^S - \hat{h}^S\|_1
\end{eqnarray}
The discriminator loss $L^S_{ACC}$, defined as the cross-entropy between the true attribute $y_{att}$ and the attribute predicted by the classifier $C^{att}$, is applied to the attribute discriminator $C^{att}$ who tries to predict the correct attribute $y_{att}$ from the latent code $\mathbf{z}^s$,
\begin{eqnarray}
\label{eq:Lacc}
L^S_{ACC} = - \log P_{W_{C^{att}}}(y_{att} | z^S)
\end{eqnarray}
where $W_{C^{att}}$ represents the weights of the classifier $C^{att}$

The adversarial loss $L^S_{ADV}$, defined as the cross-entropy between the wrong attribute $1-y_{att}$ and the one predicted by the classifier $C^{att}$, as
\begin{eqnarray}
\label{eq:Ladv}
L^S_{ADV}  = - \log P_{W_{E^{att}}}(1-y_{att} | z^S),
\end{eqnarray}
where $W_{E^{att}}$ represents the weights of the encoder $E^{att}$.

The objective of this loss is that the classifier $C^{att}$ can not predict the the correct attribute $y_{att}$ from the latent code $\mathbf{z}^s$. This is defined in order to make the latent code $\mathbf{z}^s$ independent on the variable $y_{att}$. This loss is optimized jointly with the autoencoder loss $L^S_{RC}$, then modifying the weights of the encoder $E^{att}$.
In equations (\ref{eq:Lacc}) and (\ref{eq:Ladv}), the cross-entropies are averaged over all sentences.

As described in \citep{noe2020adversarial}, a discriminator that is pre-trained on the speaker embedding $\mathbf{h}^s$ is employed to substitute the binary attribute $y_{att} \in \{0, 1\}$ by the smooth posterior probability of the discriminator $\tilde{y}_{att} \in [0,1]$.
Finally this fader is directly plugged into the speaker space of the VC system, after the speaker encoder $E^s$. We can then re-train the decoder $G^a$ of the global VC system, which we did in one of the configurations  in the experimental section.


\section{Experiments}
\label{sec:experiments}

\subsection{Dataset}

The English multi-speaker corpus VCTK 
\citep{vctk2017}  is  used for VC and gender models training and gender conversion.  The VCTK dataset contains speech data uttered by 110 speakers and the corresponding text transcripts. Each speaker read about $400$ sentences selected from English newspaper, which represents a total of about $44$ hours of speech. Two speakers were removed during experiments: one speaker (p315) for which one text file is missing (as usually removed from this dataset), and one another speaker whose name is very different from the other ones (s5). 
All speakers are included into the training and validation sets. For each speaker, we split the database in a training set with 90\% of the sentences and a validation set with 10\% of them.  The  total  duration  of the database is around $27$ hours after removing silences at the beginning and at the end of each sentence.

\begin{figure}[ht!]
\vskip 0.2in
\begin{center}
\includegraphics[width=0.9\columnwidth]{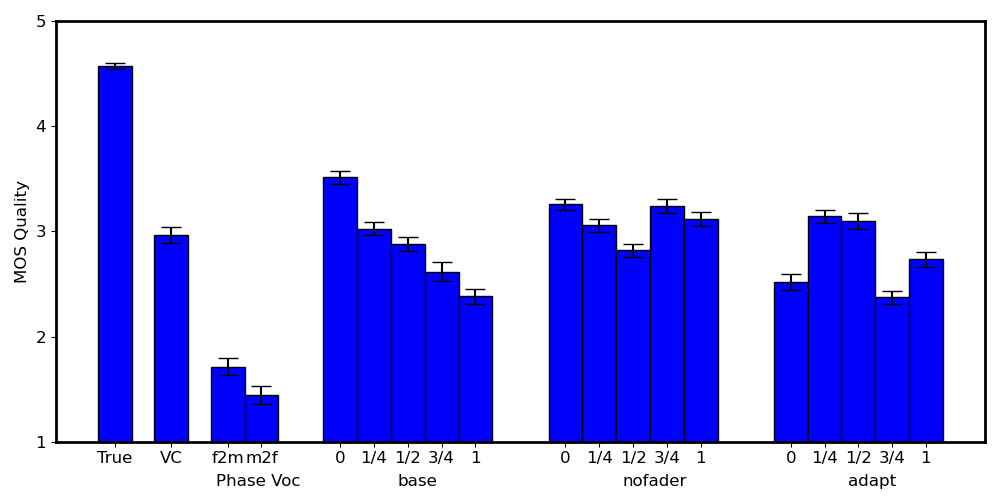}
\includegraphics[width=0.9\columnwidth]{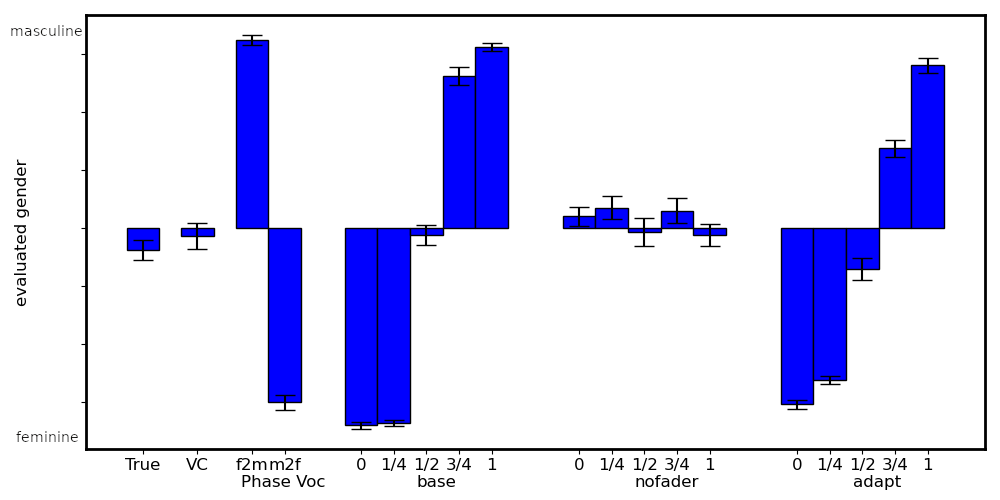}
\caption{On top: MOS score obtained for the six configurations (mean and 95\% confidence interval). On bottom:  perceived voice gender for the six configurations (mean and 95\% confidence interval).}
\label{figure:subj-eval}
\end{center}
\vskip -0.2in
\end{figure}

\subsection{Objective evaluations}

To assess whether the proposed framework is successful to disentangle speaker identity and gender representation, a set of objective evaluations were conducted: a gender classification task (including a short ablation study on the fader structure), a speaker classification task, the mutual information between the embeddings and the true gender, and a 2-D visualization of the embeddings.

\subsubsection{Gender recognition}
\begin{table}[t]
\caption{Ablation study: gender classification accuracy using the pre-trained discriminator computed on the original speaker embedding $\mathbf{h}^s$  and the reconstructed speaker embedding conditioned on the gender $w$. The dimension of the speaker embedding $\mathbf{h}^s$ is 128 and the dimension of the latent code $\mathbf{z}^s$ to 60.}
\label{table:gender_accuracy}
\vskip 0.15in
\begin{center}
\begin{sc}
\begin{tabular}{lc}
\cline{1-2}
 & gender accuracy [\%]\\
\cline{1-2} 
\textit{with adv. loss}     &         \\
Original $\mathbf{h}^s$ & 99.2\\
Est. Gender ($w=\tilde{w}$) & 99.0\\
Inv. Gender ($w=1-\tilde{w}$) & 0.8\\
De-gender ($w=1/2$) & 54.6\\
\cline{1-2} 
\textit{without adv. loss}  &  \\
Original $\mathbf{h}^s$ & 99.2\\
Est. Gender ($w=\tilde{w}$)  & 99.2\\
Inv. Gender ($w=1-\tilde{w}$)& 98.8\\
De-gender ($w=1/2$)  & 99.1\\
\cline{1-2} 
\end{tabular}
\end{sc}
\end{center}
\vskip -0.1in
\end{table}
Table \ref{table:gender_accuracy} reports the gender classification accuracy computed with the pre-trained gender discriminator at the original speaker embedding $\mathbf{h}^s$ ({\it original}) or the reconstructed speaker embedding $\widehat{\mathbf{h}}^s$ of the gender autoencoder with different values of gender conditioning $w$: with the estimated gender $\widehat{w}$ from the original speech signal ({\it est. gender}), by swapping to the opposite gender $1-\widehat{w}$ ({\it inv. gender}), or by neutralizing the gender $1/2$ ({\it de-gender}). With the adversarial setting the original speaker embedding and the reconstructed speaker embedding with the estimated gender have very high accuracies. When swapping the gender by conditioning the reconstructed embedding on the opposite of the estimated gender, the accuracy turns to zero, which is expected because the gender is inverted. With a reconstruction conditioned on $1/2$, the accuracy is around 50 \% which would correspond to a random decision in a binary classification problem. An ablation study conducted by removing the adversarial component from the fader network, the accuracies are very high in all conditions, which means that the gender conditionning is ineffective. Therefore the adversarial loss is needed to disentangle the speaker gender from the rest pf the speaker identity. This shows that the adversarial loss is required and highly efficient to disentangle and manipulate speaker gender from speaker identity.

\subsubsection{Speaker classification}
\begin{figure}[h]
\vskip -0.05in
\begin{center}
\centerline{\includegraphics[width=0.6\columnwidth]{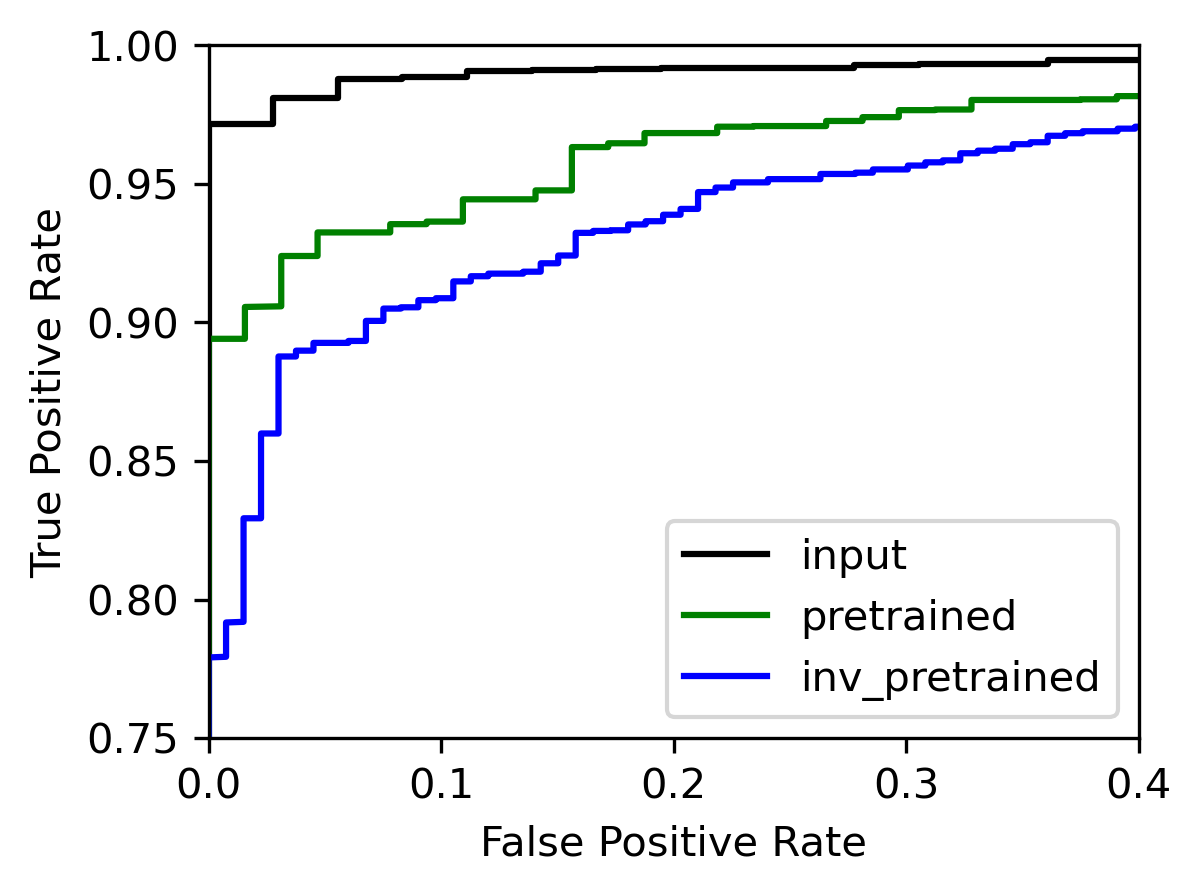}}
\caption{ Receiver Operating Characteristic curves of speaker classification using speaker encoder classifier as computed on the original speaker embedding $\mathbf{h}^s$  and the reconstructed speaker embedding conditioned on the gender $w$}
\label{figure:roc}
\end{center}
\end{figure}

\begin{table}[t]
\caption{Equal Error Rates in percentages of speaker classification using speaker encoder classifier computed on the original speaker embedding $\mathbf{h}^s$  and the reconstructed speaker embedding conditioned on the gender $w$. The dimension of the speaker embedding $\mathbf{h}^s$ is 128 and the dimension of the latent code $\mathbf{z}^s$ to 60.}
\label{table:speaker_eer}
\vskip 0.15in
\begin{center}
\begin{sc}
\begin{tabular}{lc}
\cline{1-2} 
& EER [\%]\\
\cline{1-2} 
Original $\mathbf{h}^s$ & 2.8\\
Est. Gender ($w=\tilde{w}$) & 6.9\\
Inv. Gender ($w=1-\tilde{w}$) & 9.2\\
De-gender ($w=1/2$)  & 6.8\\
\cline{1-2} 
\end{tabular}
\end{sc}
\end{center}
\vskip -0.1in
\end{table}
Figure \ref{figure:roc} presents the Receiver Operation Characteristics (ROC) curves corresponding to the speaker classification from the original speaker embedding and the reconstructed speaker embedding conditioned on the gender and Table \ref{table:speaker_eer} summarizes the equal error rates (EERs) obtained from the original speaker embedding and the reconstructed speaker embedding conditioned on the gender. The EER is very low (2.8 \%) for the original speaker embedding, which indicates that the speaker classifier is very efficient to determine the speaker identity from the speaker embedding. The manipulation of the gender conditioning $w$ degrades the EER in all cases, but these rates remain relatively low:
around 6.8 \% for the gender estimated from the pre-train classifier and for $w=1/2$, and 9.2 \% when the gender is swapped. This means that most of the speaker identity is preserved after gender manipulation. However, the speaker identity cannot be totally preserved since identity and gender are certainly not linearly separable variables.

\begin{figure}
\begin{center}
\includegraphics[width=0.45\columnwidth]{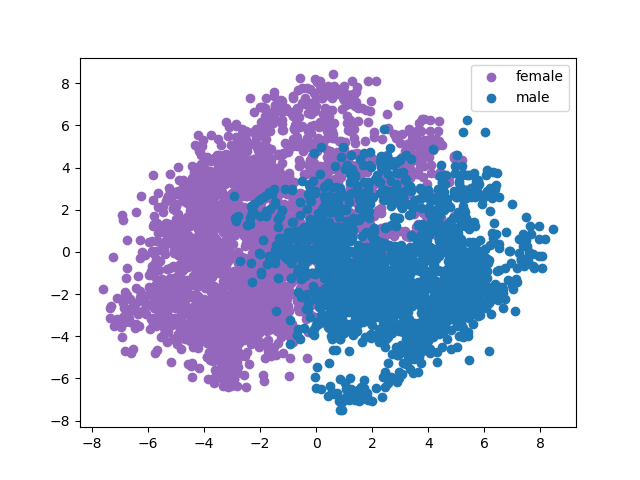}
\includegraphics[width=0.45\columnwidth]{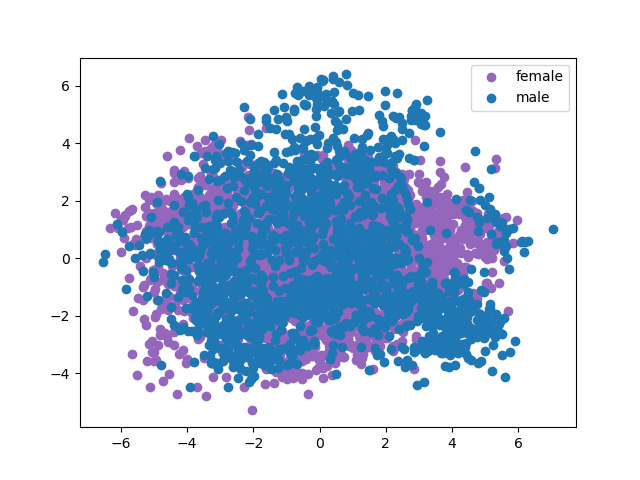}
\caption{On top: PCA visualization of the speaker encoder embeddings $\mathbf{h}^s$ on the evaluation set. The selected components are 1 and 2. On bottom: PCA visualization of the latent code $\mathbf{z}^s$ on the evaluation set. The selected components are 3 and 7.}  
\label{figure:pca_input}
\end{center}
\vskip -0.2in
\end{figure}

{
\subsubsection{Mutual Information and Embeddings visualization}
\begin{table}[t]
\caption{Approximation of the mutual information between the true gender and the continuous multi-dimensional embedding. The dimension of the speaker embedding $\mathbf{h}^s$ is 128 and the dimension of the latent code $\mathbf{z}^s$ to 60.}
\label{table:mutual_info}
\vskip 0.15in
\begin{center}
\begin{sc}
\begin{tabular}{lc}
\cline{1-2} 
& Mutual Information \\
\cline{1-2} 
Original $\mathbf{h}^s$ & 0.47\\
Est. Gender ($w=\tilde{w}$)& 0.44\\
Inv. Gender ($w=1-\tilde{w}$)& 0.38\\
De-gender ($w=1/2$)  & 0.16\\
Latent code $\mathbf{z}^s$ &  0.11\\
\cline{1-2} 
\end{tabular}
\end{sc}
\end{center}
\vskip -0.1in
\end{table}

Table \ref{table:mutual_info} present the approximated calculation of the mutual information between the true gender and the original speaker embedding and the conditionally reconstructed speaker embeddings. The score is computed using an estimator of the mutual information between discrete and continuous variables as described in \citep{Gao_MI_2017}. The dimension of the continuous data is reduced from 128 to 8 using PCA and the mutual information is obtained by selecting the pair of coordinates that maximize the latter.
The PCA coordinates used to plot the 2-D visualizations in Figure \ref{figure:pca_input} are selected in the same way. 
From Table \ref{table:mutual_info}, the mutual information corresponding to the latent code $\mathbf{z}^s$ and the de-gender $w=1/2$ are much lower that the others. This indicates that the latent code $\mathbf{z}^s$ contains very little information about the gender and becomes mostly gender-independent as illustrated in Figure \ref{figure:pca_input}, and that the conditioning $w=1/2$ successfully generates a speaker embedding that is also mostly genderless.  
This highlights the fact that the disentanglement between speaker identity and gender is highly effective.

}

\subsection{Subjective Evaluation}

To assess whether the proposed architecture is efficient to convert the gender of the voice, a subjective evaluation was conducted. The task consisted into listening one speech sample (converted or not), and to judge: 1) whether the voice is typically perceived as: {\it feminine}, {\it rather feminine}, {\it uncertain}, {\it rather masculine}, or {\it masculine}; 2) the sound quality on a standard Mean Opinion Score (MOS) 5-degree scale from 1 ({\it bad}) to 5 ({\it perfect}) which is commonly used for the experimental evaluation of Text-To-Speech and Voice Conversion systems. Each participants had to judge 20 speech samples which were randomly selected among the total of speech samples produced for the subjective experiments.
Four speakers were used for the experiment: two males (p232 and p274) and two females (p253 and p300) with five randomly chosen sentences per speaker in the validation set.
Six configurations were compared: 1) the original audio signal ({\it true}) and the converted audio signal with~: 2) the original VC system ({\it VC}); 3) a phase vocoder  ({\it phase voc.}, see supplementary for details), with two cases~: female-to-male conversion ({f2m}) and male-to-female conversion ({m2f}); 4) the VC system with the proposed gender autoencoder ({\it base}) with five conditionning values of the parameter $\tilde{w} \in \{0, 1/4, 1/2, 3/4, 1\}$; 5) 
the VC system with the gender autoencoder {but trained without the fader loss} ({\it nofader}), with the five values of the parameter $\tilde{w}$; and 6)
 the VC system with the gender autoencoder with the VC decoder re-trained ({\it adapt}) with the five values of the parameter $\tilde{w}$. All neural VC systems are implemented with the time-synchronous version presented in this paper, and the phase vocoder is also time-synchronous so that the timing of the original speech samples and all converted speech samples are time-synchronous.
 
Figure \ref{figure:subj-eval} presents the MOS scores and the perceived gender for the compared system configurations (mean and 95\% confidence interval). Regarding the perceived quality: the original speech samples has the higher scores ($4.6$), the original VC system has similar scores as the ones reported in \citep{Zhang_2020} ($2.90$), and the speech converted with the phase vocoder has pretty low scores (1.6) which is due to the use of the default settings and indicate the limitation of voice conversion based on signal processing only. The three versions of our proposed VC system have similar scores that are comparable to the one of the original system (between $3.0$ and $4.0$): $2.9$ for the {\it base} VC system, $3.11$ for the {\it nofader} VC system, and $2.97$ for the {\it adapt} VC system. This shows that the extra-add of the gender auto-encoder does not degrade the conversion quality. This quality tends to degrade in the case of the {\it base} VC system from female to male, but this trends tend to disappear for the{\it adapt} VC system in which the VC decoder is re-trained together with the gender auto-encoder. Regarding the perceived gender, the true gender is well-recognized for the original speech samples, the converted speech with the original VC system, and the converted speech with the phase vocoder. As mentionned previously, the VC system with a gender autoencoder without the fader loss is totally inefficient to convert the gender. For the proposed VC system with the gender auto-encoder, the gender conditionning is efficient to manipulate the perceived gender during conversion as one can observe a clear variation of the perceived gender with respect to the conditionned gender. In the {\it base} VC system, there is however a discontinuity around the value $w=1/2$ which means that the conversion jumps from female to male and fails to generate genderless voices. This appears to be much more linear in the {\it adapt} VC system, which indicates again that the re-training of the VC decoder can also improve the conversion around the genderless value ($w=1/2$).

\section{Conclusion}
\label{sec:conclusion}

This paper presents a structured neural VC architecture that allows the manipulation of voice attributes (e.g., gender and age) based on the adversarial learning of a hierarchically structured speech and speaker encoding. The proposed VC architecture employs  multiple auto-encoders are used to encode speech as a set of idealistically independent  linguistic and extra-linguistic representations, which are learned adversarially and can be manipulated during VC. Moreover, the proposed architecture is time-synchronized so that the original voice timing is preserved during conversion which allows lip-sync applications.
A set of objective and subjective evaluations conducted on the VCTK dataset shows the efficiency of the proposed framework on the task on voice gender manipulation. Further work will investigate the generalization of the proposed framework to other voice attributes, such as age, attitudes, and emotions.


\bibliographystyle{natbib}
\bibliography{example_paper}

\clearpage
\appendix

\section*{Supplementary material}

\vspace{0.25cm}

\section{Implementation details}

\subsection{S2S VC architecture}

The model configuration parameters are the same as the ones described in \citep{Zhang_2020}, with exception to the recognition encoder $\mathbf{E}^r$ and the decoder $\mathbf{G}^a$ (referred to as $\mathbf{D}^a$ in \citep{Zhang_2020}) that are modified for the time-synchronized VC system. 
Table \ref{tab:VC_config} presents the details of these modification, together with the components of the fader network used for the identity and gender disentanglement: the encoder $\mathbf{E}^{att}$, the classifier $\mathbf{C}^{att}$, and the decoder $\mathbf{G}^{att}$. The simplifications realized to the recognition encoder $\mathbf{E}^r$ and the decoder $\mathbf{G}^a$ enable time-synchronous conversions and a consequent saving in computational time: approximately 33\% of the computational time for training on our server with a single GPU GForce GFX 1080Ti and on the Jean-Zay super-computer using a single GPU (Tesla V100-SXM2-32GB) \footnote{This work was performed using HPC resources from GENCI-IDRIS (Grant 2020-[AD011011757])}.

\begin{table}[h!]
\caption{Details of the model configuration. FC refers to a fully-connected layer, BLSTM to a bi-directional LSTM, and Tanh to the hyperbolic tangent activation function.}
\label{tab:VC_config}
\small
\vspace{0.25cm}
\centering
\begin{tabular}{p{1cm}p{5cm}}
\cline{1-2} 
\multirow{4}{*}{$\mathbf{E}^r$} 
 & \\
 & 2 layers BLSTM-Dropout(0.2), 256 cells each direction $\rightarrow$  \\
 & FC-512-Tanh \\ 
 & \\
\cline{1-2} 
\multirow{5}{*}{$\mathbf{E}^s$}
 & \\
 & 2 layers BLSTM-Dropout(0.2), 128 cells each direction $\rightarrow$  \\
 & average pooling  $\rightarrow$ \\ 
 & FC-128-Tanh  \\
 & \\
\cline{1-2} 
\multirow{4}{*}{$\mathbf{G}^a$} 
& \\
& 2 layers BLSTM, 64 cells each direction $\rightarrow$  \\
&  FC-80 \\ 
& \\
\cline{1-2} 
\multirow{3}{*}{$\mathbf{E}^{att}$} 
& \\
& FC-60  \\
& \\
\cline{1-2} 
\multirow{3}{*}{$\mathbf{G}^{att}$} 
& \\
& FC-1  \\
& \\
\cline{1-2} 
\multirow{3}{*}{$\mathbf{G}^{att}$} 
& \\
& FC-128-Tanh  \\
& \\
\cline{1-2} 
\end{tabular}
\end{table}

\subsection{Pre- and post-processing}

Following \citep{Zhang_2020} the system operates on a mel spectrogram representation of the speech signal. For the signal analysis we follow the parameterization proposed in \citep{liu_wavenet_2018}, that is the input signal is down-sampled to 16kHz, converted into an STFT using an Hanning window of $50$ms with hop size of $12.5$ms and an FFT size of $2048$. We then use $80$ Mel bins covering the frequency band from $0$ to $8$Khz and convert the result into log amplitude domain. 
A standardization of the log-mel-spectrogram is applied at the input of the VC system, i.e. on each Mel bin, removing the mean and diving by the standard deviation, which are pre-computed on the entire training dataset.
For rendering audio from a generated mel spectrogram a multi speaker approach is required because the generated mel spectrograms are not linked to any existing speaker identity. We initially used a Griffin and Lim \citep{Griffin/Lim:84} algorithm for phase reconstruction, which however did not provide sufficient quality for perceptual evaluations. We then resorted to a multi-speaker waveglow type decoder following loosely \citep{prenger2018waveglow}. This decoder was trained over $900,000$ iterations using all samples of the VCTK database with batch size $50$, and segment length $375$ms using the Adam optimizer with learning rate $1e^{-4}$. While the quality of this decoder is far from perfect it is consistently better than the quality obtained with the Griffin and Lim algorithm and is therefore used for the perceptual tests. The decoder has a slight tendency to produce a overly rough voice quality indicating an instability on the stability of the F0. The decoder is subject to further research and will be published elsewhere. The trained network together with the script used for inversion will be made available as part of the supplementary material.

\subsection{Computation infrastructure and runtime costs}

All training has been run on a single GPU (GForce GFX 1080Ti). The inference and the mel inversion have been run on the CPU (Xeon(R) CPU E5-2630 v4 @ 2.20GHz), while some of the figures have been generated using the GPU. 
The duration of the VC model training is 20 minutes per epoch with 80 epochs (roughly 27 hours) and the training of the gender autoencoder model lasts 1 minute and 30 seconds per epoch with 400 epochs (total of 10 hours). 
The inference of  one sentence of 1.5 seconds takes around 2 seconds for computing the mel-spectrogram  plus two seconds for the mel inversion, using our CPU.

Concerning the training parameters, the VC system makes use Adam optimizer with a learning rate equal to $1e^{-3}$ and the batch size is 32, while the training of the gender autoencoder is done with the SGD optimizer with a learning rate equal to $1e^{-4}$ and the momentums set to 0.9, with a batch size equal to 64.
Also the pre-trained gender discriminator makes use of the SGD optimizer also with a learning rate equal to $1e^{-4}$ and the momentums equal to 0.9, three epochs are used each epoch lasting 1 minute and 30 seconds, with a batch size equal to 64.

\subsection{Hyper-parameter setting of the network}

Concerning our synchronous version of the Voice Conversion network, we have modified the recognition encoder $E^r$ and the decoder $G^a$, because they were originally auto-regressive.
The recognition encoder and the decoder are both composed of two bi-directional LSTMs and a Fully Connected layer, with a Tanh activation function in $E^r$ (see table \ref{tab:VC_config} for more details). 
We tried to increase the complexity of the decoder by adding up to four layers but informal listening of voice conversions on the training data led us to abandon this option.

For the gender autoencoder, our starting point is the paper \citep{noe2020adversarial}. We have removed the length normalization and replaced the loss based on the cosine distance by the mean absolute error between the input and output of the autoencoder. 
The choices on the paper \citep{noe2020adversarial} are specific to Automatic Speaker Verification. 
We kept the standardization on the input of the autoencoder, with the converse operation at the output.

Concerning the dimension of the latent code of the gender autoencoder $z^s$, we chose the value 60 after informal listening of converted audio signals on the training database (with gender dimensions varying between 40 and 100). 
We believe that the chosen value is a compromise between the bottleneck effect produced by the auto-encoder and its complexity in terms of trainable parameters.

\section{Resources for reproduction of results}

Our code is implemented in Python 3.7 with TensorFlow 2.2. 

The code and results will be made available in from of a github repository after publication. 


The resources made available  cover: 

\begin{itemize} 
\item the network models and scripts used for training and inference of the gender conversion network,
\item pre-trained network weights that have been used for the perceptual tests
\item the waveglow style network with weights and a script that allows running the mel to audio inversion.
\end{itemize}

All data-sets are public and will therefore not be included into the repos.

\section{Gender conversion baseline}

To the best of our knowledge there are no  neural gender conversion algorithms available in the literature, and therefore we use a traditional signal processing approach as our baseline for perceptual tests.


\begin{figure}[!ht]
\begin{center}
{\includegraphics[width=0.45\columnwidth]{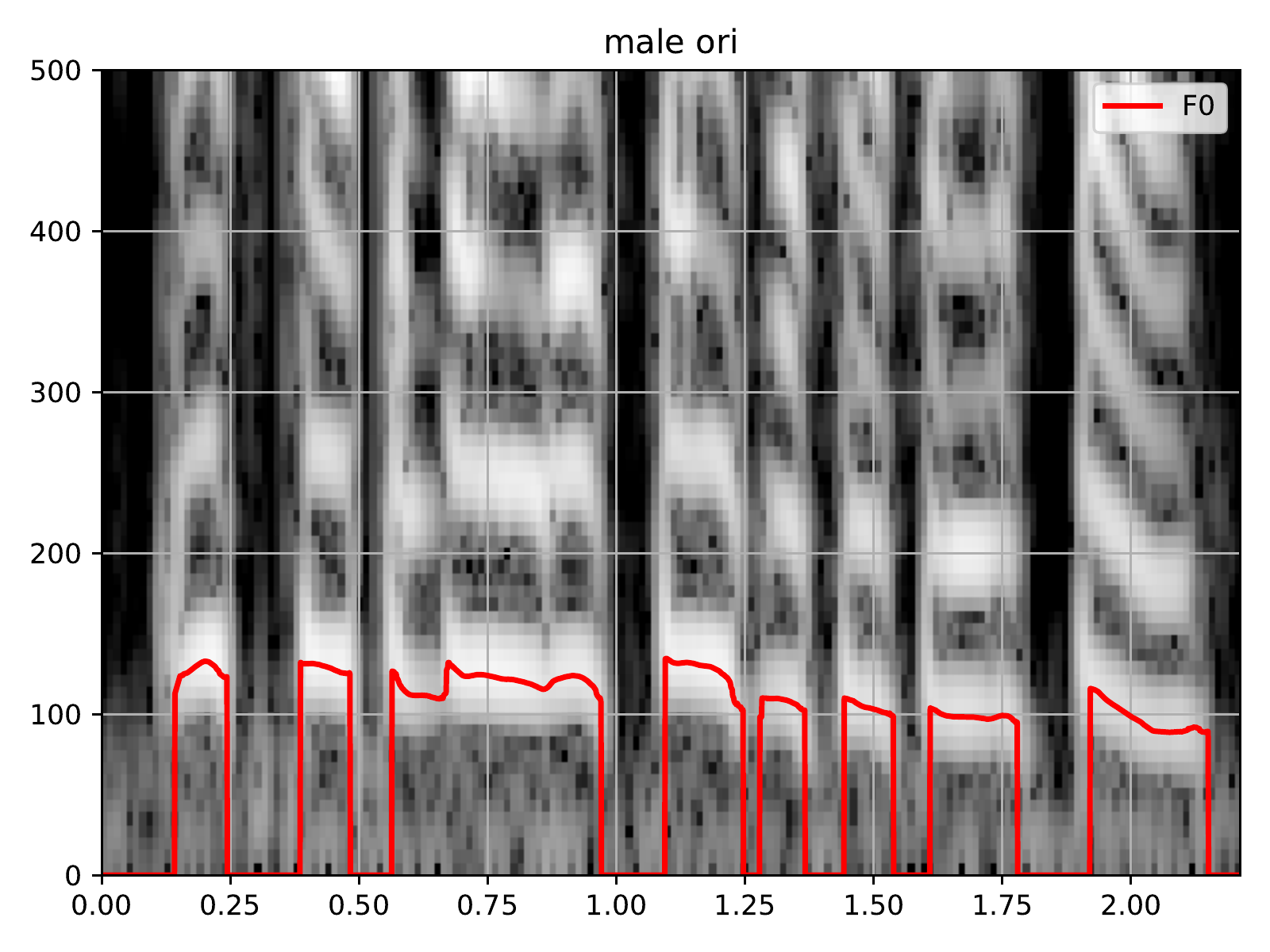}}
\hspace{0.1cm}
{\includegraphics[width=0.45\columnwidth]{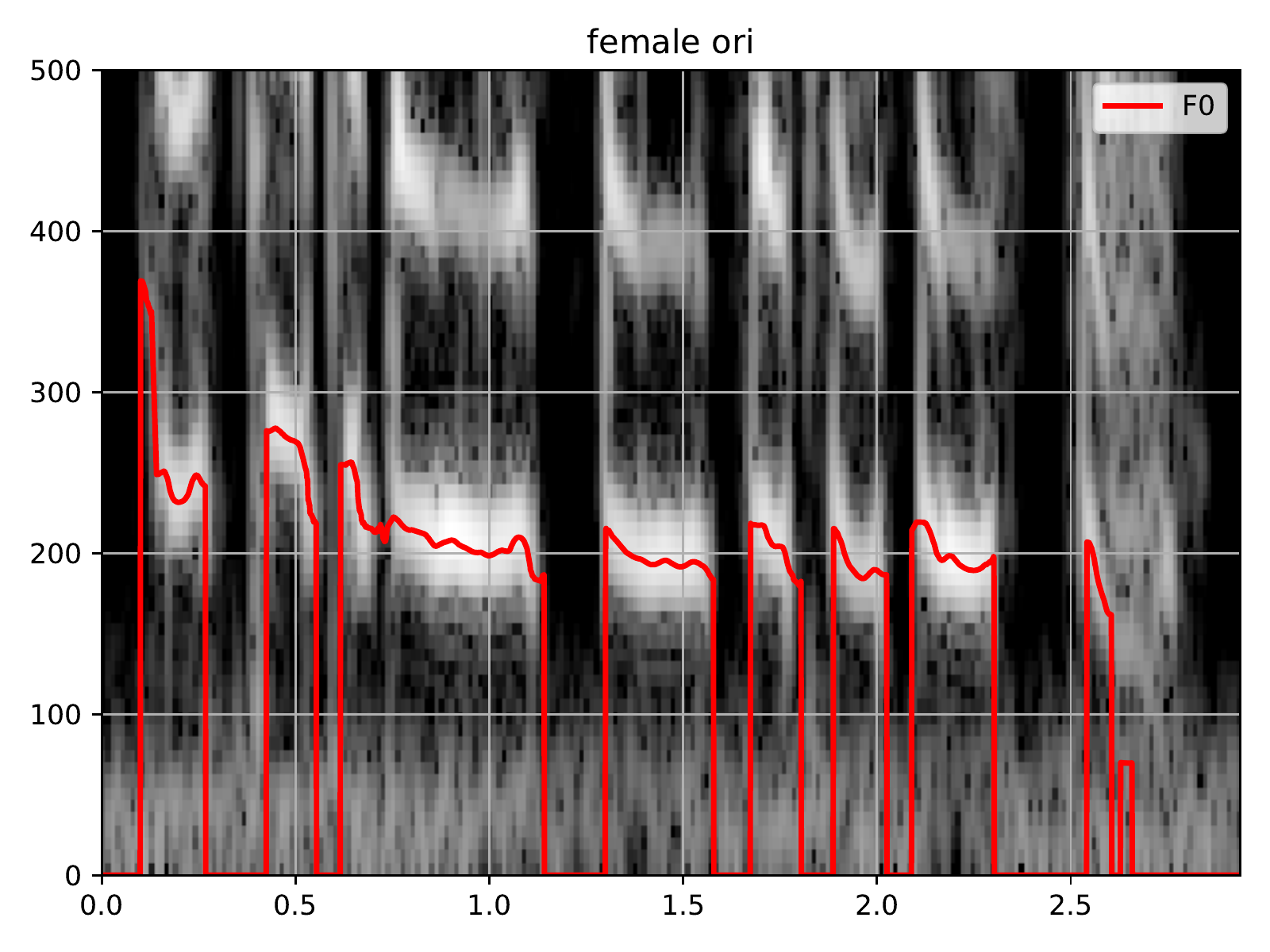}}\\
{\includegraphics[width=0.45\columnwidth]{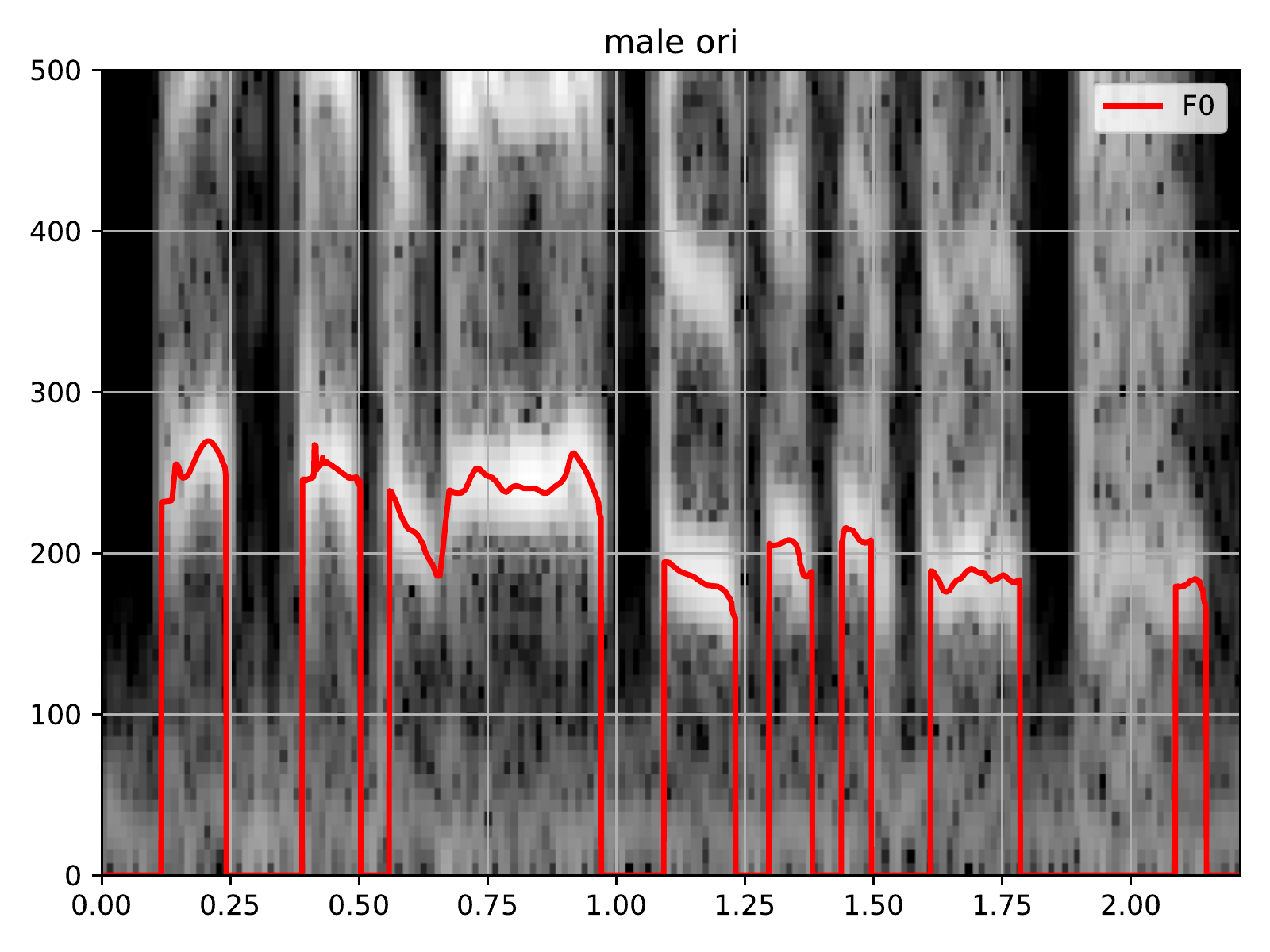}}
\hspace{0.1cm}
{\includegraphics[width=0.45\columnwidth]{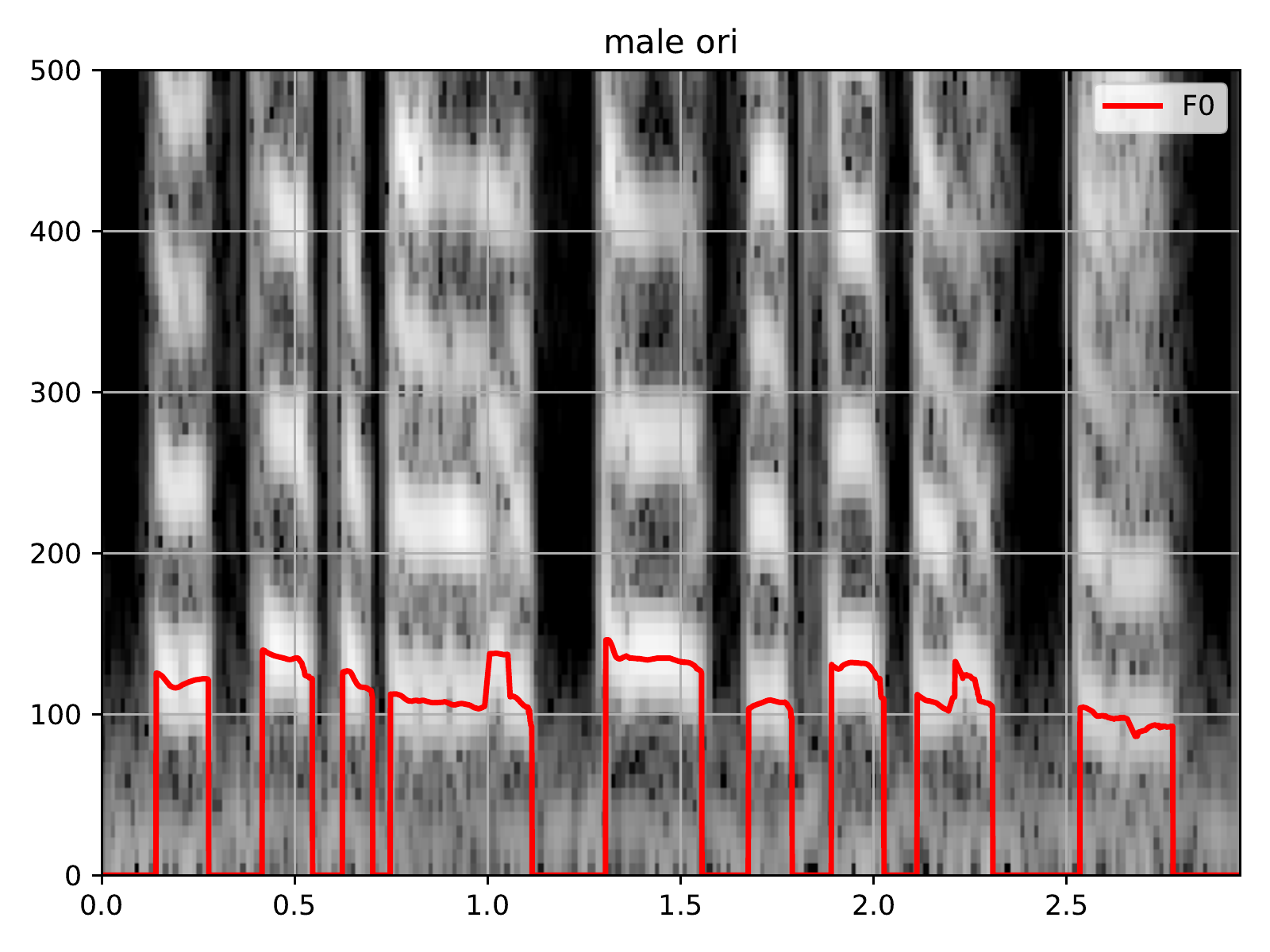}}
\caption{Visualization of the spectrograms and the F0 curves of the sentence "Ask her to bring these things with her from the store". 
On the (left) images, the sentence is uttered by a male speaker and on the (right) images the sentence is uttered by a female speaker. On the (top),  the original audios and on the (bottom) the gender converted audios. The solid red line is the F0. On y-axis: frequency in Hertz; on x-axis: time in seconds.}
\label{figure:pitch_visualization}
\end{center}
\end{figure}


Classic voice transformation algorithms perform gender manipulation by means of modifying in average the fundamental frequency (F0) and the positions of the vocal tract resonances (also known as formants). Due to physiological differences between the female and male voice organs (notably the size of vocal folds and vocal tract)  these two parameters have average values which generally differ for male and female voices. The differences have been measured and documented in the literature \citep{Peterson:52, Iseli:07}. Given that these parameters are part of the physiological configurations of the vocal organs they are part of the speaker identity and it has been shown in  \citep{Farner_2009} that a constant but independent transposition of the F0 and the formants can be used to successfully modify the perceived gender and age of a voice.
Following these findings we use the following parameters for gender conversion: F0 is shifted by  $\pm$ one octave ($\pm$1200 cents) and the spectral envelope is shifted by $\pm$ 3 semi-tones (i.e., $\pm$300 cents) where the sign of the shift depends  on the gender of the original sound. For male to female the positive sign is used, and the negative sign for female to male conversion.  A shape invariant phase vocoder \citep{Roebel:10b} is used for the signal manipulation, by using the true envelope estimator for the representation of the formant structure \citep{Roebel/Rodet:05a}. These types of algorithms have been used successfully in the past for gender transformation for professional uses.
However, the default setup does not work equally well for all voices and manual fine tuning is generally employed to optimize the coherence of the transformed voice signal. As the proposed algorithm is fully automatic we don't apply manual tuning for the signals used in the subjective tests.

\section{Pitch conversion visualization}


Figure \ref{figure:pitch_visualization} shows four spectrograms superimposed with related pitch contours (F0, in red solid lines). The sentence "{\it Ask her to bring these things with her from the store}" is uttered by a male speaker (left) and by a female speaker (right). The (top) figures show the original signals and the (bottom) figures correspond to the conversion conditioned on the opposite gender. 
The gender conversion algorithm clearly transposes the average F0 as has been used for the classical algorithm ($\pm$ 1 octave). Moreover, one can also see that this transposition is clearly dynamic, then affecting the whole intonation when converting from a masculine to a feminine voice (and vice versa). Additionally, the algorithm created vocal fry at the final words of the utterance when converting from male to female, while this is the opposite when  converting from a female to a male voice. We conjecture that this presence or absence of vocal fry reflects a general tendency of the male and female voices in the database. 

\section{Definition of the specific measure or statistics used to report results}
\begin{itemize}
    \item A \textbf{Receiver Operating Characteristic curve}, or ROC curve, is a graphical plot that illustrates the diagnostic ability of a binary classifier as its discrimination threshold is varied. The ROC curve is created by plotting the true positive rate (TPR) against the false positive rate (FPR) at various threshold settings.
    \item The \textbf{Equal Error Rate} (EER) is the error rate of a binary classifier when the operating threshold for the accept/reject decision is adjusted such that the probability of false acceptance and that of false rejection become equal. On the ROC curve, it corresponds to the intersection with the anti-diagonal line.
    \item The definition of the \textbf{Mutual Information} (MI) between two continuous random variables $X, Y$ is equal to $\int_{\mathbb{R}} \int_{\mathbb{R}} \log \frac{p_{X,Y}(x, y)}{p_X(x) p_Y(y)} p_{X,Y}(x, y) dx dy$. The Mutual Information is a measure of the mutual dependence between the two variables. More specifically, it quantifies the "amount of information"  obtained about one random variable through observing the other random variable. In our paper, we measure the Mutual Information between a discrete variable (the gender) and a two dimensional continuous variable. It is estimated using  k nearest neighbors approaches \citep{Gao_MI_2017}. 
    \item In our context, the \textbf{Mean Opinion Score} (MOS) is a measure representing overall \textit{perceived quality} of the audios generated by a given system. It is the arithmetic mean over all individual values on a range from 1 to 5 that the subjects assign to their opinions of the quality of an audio,  1 being lowest perceived quality, and 5 being the highest perceived quality.
In the subjective evaluation, for the quality MOS  and the perceived voice gender, we plot confidence intervals using plus or minus the standard deviation of the mean quantities multiplied by 1.96 (which is a 95 \% confidence interval under Gaussian assumption).
Note that the standard deviation of the mean of $n$ values is equal to standard deviation of all values divided by the square root of $n$.
\end{itemize}

\end{document}